# Advancements in Complementary Carbon Nanotube Field-Effect Transistors

Ali Javey, Qian Wang, Woong Kim, and Hongjie Dai*
Department of Chemistry and Laboratory for Advanced Materials, Stanford University, Stanford, CA 94305, USA.
*Tel: (650) 723-4518, Fax: (650) 725-9793, E-mail: hdai@stanford.edu

**Abstract**
High performance p- and n-type single-walled carbon nanotube (SWNT) field-effect transistors (FETs) are obtained by using high and low work function metals, Pd and Al as source/drain (S/D) electrodes respectively. Ohmic contacts made to chemically intrinsic SWNTs, with no or small Schottky barriers (SB), afford high ON-state currents up to 20 µA per tube. The lack of significant Fermi-level pinning at the nanotube-metal interfaces allows for fine-tuning of the barrier heights for p-and n-channel conductions by changing the contact metals. The air-stable p- and n-FETs thus obtained can be used for complementary nanoelectronics, as demonstrated with the fabrication of an inverter. Other important issues regarding nanotube FETs including hysteresis, OFF-state leak currents, choice of nanotube diameter, and threshold voltage control are discussed.

## Introduction

There has been much progress recently in the fabrication, understanding and exploring the performance limits of SWNT-FETs (1-7). Much effort is still needed to address some of the issues key to potential CMOS applications of nanotubes, including the strategies to fabricate stable and robust complementary FETs, formation of reliable ohmic contacts for high current delivery capability, and transistor scaling effects such as OFF-state leakage. The current work addresses some aspects of these issues and presents the highest performance p- and n-type SWNT-FETs made to date by applying the appropriate S/D contact metals.

## Device Fabrication

The SWNT devices used for this work were obtained by patterned chemical vapor deposition (CVD) of SWNTs on Si/SiO$_2$ substrates followed by metal S/D contact formation (8). Degenerately p-doped Si was used as the bottom gate and thermal SiO$_2$ ($t_{ox}$ ~ 10 nm) was used as the gate dielectric (Fig. 1). For all of the devices, the S/D distance, or the channel length, was L ~ 3 µm, the devices were passivated by PMMA (9), and the individual SWNTs in the FETs had diameters in the range of d ~ 2 to 3 nm (band gap E$_g$ ~ 0.45 to 0.3 eV). It is important to note that tube diameter is critical to transport and contact properties of SWNT-FETs since E$_g$ scales with 1/d.

## p-type FETs with Pd Ohmic Contacts

We have taken systematic efforts to investigate various metals for S/D electrodes (top-contact) in order to optimize the contacts for SWNT-FETs. As recently reported (1), among the other metals tested, including Pd, Ti, Ti/Au

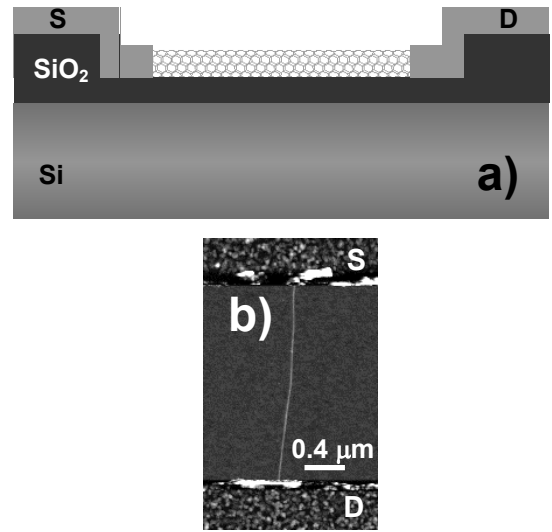

Fig. 1. (a) Schematic side view of a back-gated SWNT-FET with 10 nm thick SiO$_2$ gate dielectric under the channel and 100 nm SiO$_2$ under the bonding pads. (b) An atomic force microscopy image of a device with an individual SWNT bridging the Al source (S) and drain (D) electrodes

bilayer, Ni, Mo, W and Pt, Pd is found to form the most transparent contact for p-type FETs. With Pd, ohmic contacts to the valence band of SWNTs, free of any positive SBs (Fig. 2a inset) can be made for chemically intrinsic SWNTs with d > 2 nm, owing to its high work function ($\phi_{Pd}$ ~ 5.1 eV) and favorable wetting interactions with nanotubes. Long channel devices consistently show p-channel ON-state conductance of G$_{ON}$ ~ 0.1-0.2×4e$^2$/h (R$_{ON}$ ~ 50 kΩ), ON and OFF ratio of G$_{ON}$/G$_{OFF}$ ~ 10$^5$ (for $t_{ox}$=10 nm) and subthreshold swing S ~ 100-150 mV/decade (Fig. 2). The MOSFET square law model fits well the device output characteristics for I$_{DS}$ below ~ 10 µA (or the onset of optical phonon scattering) with a hole mobility of µ$_p$ ~ 3400 cm$^2$/V.s (Fig. 2b). This result suggests that ohmically contacted p-type SWNT FETs can operate essentially like MOSFETs, at least in the ON and subthreshold regimes. Transport in L ~ 3 µm long SWNTs is typically diffusive, as indicated by the finite albeit high mobility. With Pd ohmic contacts, the saturation current can be pushed to ~ 20 µA (~ 8000 µA/µm, normalized by d), which is 5-6 times higher than that of the SWNT SB-FETs.

## n-type FETs with Near-Ohmic Al Contacts

Air-stable ohmically contacted n-type SWNT-FETs are highly desired. Contacting SWNTs with d = 2-3 nm by Al metal electrodes affords such FETs. The devices exhibit n-channel conductance of G$_{ON}$ ~ 0.05×4e$^2$/h (R$_{ON}$ ~ 100 kΩ), subthreshold swing of S ~ 150 mV/decade and $I_{ON}/I_{OFF} > 10^4$ (Fig. 4). Square law fitting (Fig. 4b) yields an electron mobility of $\mu_n$ ~ 3750 cm$^2$/Vs. The maximum saturation current for the n-FETs is ~ 10 μA (~3600 μA/μm), the highest reported to date for SWNT n-FETs, but not as high as the p-FETs with Pd contacts. This suggests slight SBs to the conduction bands of SWNTs may still exist with Al contacts. The existence of small SB in the n-FET is also evidenced in the temperature dependent characteristics of the device (Fig. 5). The conductance of the device exhibits monotonic decreases at lower temperatures, suggesting a thermionic component in the current transported through the SB. At 2 K, the device exhibits clear Coulomb oscillations in the ON-state (Fig. 5 inset), corresponding to a nanotube quantum dot with two SBs at the contacts as tunnel barriers. This is different than the SB-free p-FETs with transparent Pd contact that exhibit phase coherent Fabry-Perot interference at low temperatures for the ON-state (1). In the future, SB-free n-FETs may be achievable by applying lower work function S/D metal contacts or by chemically n-doping the nanotubes.

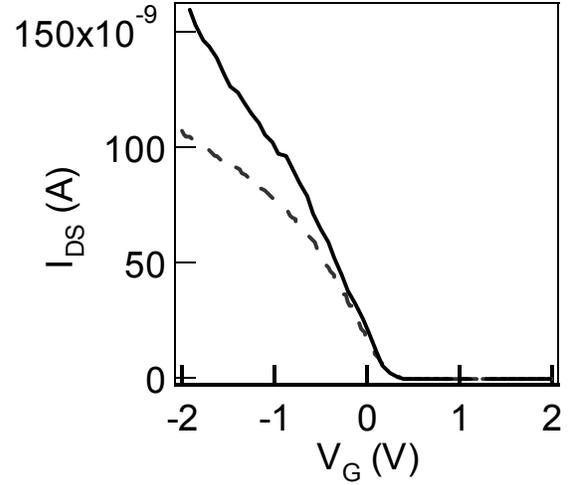

Fig. 3. Transfer characteristics of the device shown in Figure 2 before (dashed) and after (solid) PMMA passivation. $V_{DS}$ = 10 mV.

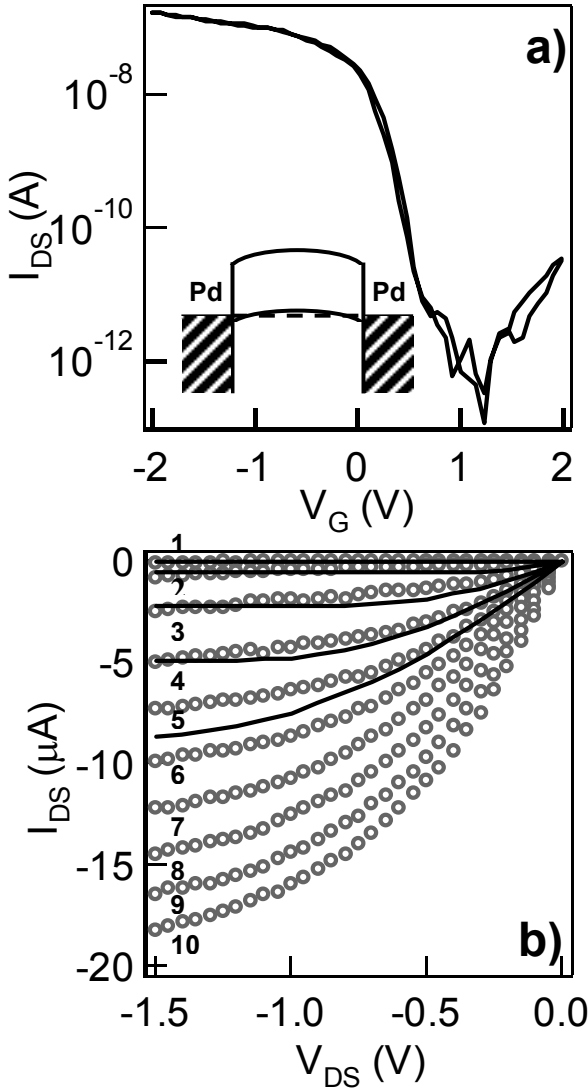

Fig. 2. Electrical properties of a Pd contacted p-FET. (a) Transfer characteristic of a device (d ~ 2.5 nm and L ~ 2.5 μm) taken with $V_{DS}$ = 10 mV after PMMA coating. The two curves represent different $V_G$ sweep directions, showing minimal hysteresis after passivation. The inset shows the band diagram for the ON state. (b) $I_{DS}$-$V_{DS}$ curves of the same device taken at different back-gate voltages. Curves 1-10 correspond to $V_G$ = 0.4 V to –3.2 V in –0.4 V steps. The solid lines are calculated from the square law model for a diffusive channel to fit curves 1-5. A deviation from the MOSFET behavior is observed for $V_G$ < -0.4 V.

By simply changing the S/D contact metals, one obtains p- or n- SWNT-FETs without relying on chemical doping of the channel. This is a distinct feature of SWNTs owing to their unique 1-D structure with fully saturated surface bonds and no interface states, resulting in no Fermi-level pinning. The high and low work functions of Pd and Al (~ 5.1 eV and ~ 4.1 eV) afford zero and small SBs to the valence and conduction bands of SWNTs respectively. At the Pd-SWNT junction, no SBs exist for p-channel conduction (large SB for n-channel, Fig. 2a inset), which is responsible for the high current delivery capability (25 μA is the optical phonon scattering limit for SWNT) and MOSFET-like operation. At the Al-SWNT contacts, small SBs exist which limits the n-channel current to ~ 10 μA (large SB to the p-channel, Fig. 4a inset). Due to the existence of small SBs, $\mu_n$ ~ 3750 cm$^2$/Vs is only a rough estimate for the lower limit of the electron mobility.

It is found that the contact metal work function is only a phenomenological parameter, useful to a limited extent in guiding the choice of metals for ohmic contacts to SWNTs. It simply does not hold true that higher work function metals give lower SBs for p-type nanotube FETs. This indicates that other important factors may also play a role, including metal-SWNT or metal-SiO$_2$ interfacial binding interactions and perhaps also the chemical environments, such as ambient oxygen and water interactions with the contacts.

## Importance of Hysteresis Elimination

The SWNT-FETs presented here, exhibit no appreciable hysteresis in the current vs. gate voltage sweeps due to PMMA passivation to remove surface water molecules (9), leading to believable carrier mobilities ($\mu_h \sim 3400$ cm$^2$/V.s; $\mu_n \sim 3750$ cm$^2$/V.s). Mobility values extracted for devices without passivation can be up to 20,000 cm$^2$/V.s depending on the branch of the hysteretic $I_{DS}$-$V_{GS}$ curves used for the analysis. Elimination of hysteresis is, therefore, essential to understanding the intrinsic transport properties of semiconducting SWNT-FETs, as illustrated in detail earlier (9). Mobilities reported for nanotubes are credible only if surface passivation is used to remove charges induced by simple chemical environment factors. Furthermore, the operating threshold voltage of nanotube FETs is precisely extracted only when the un-controlled hysteresis effects are eliminated.

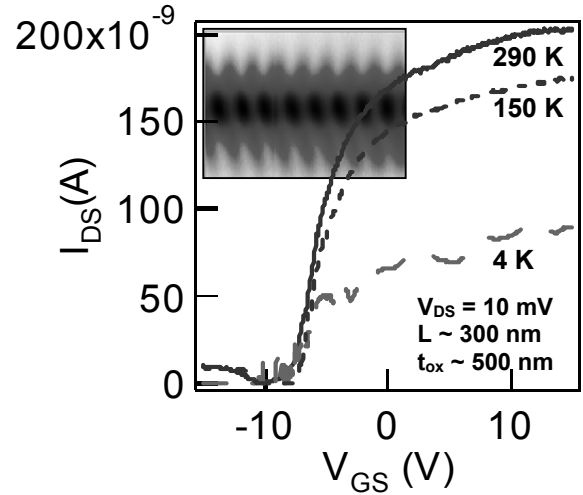

Fig. 5: Temperature dependent characteristics of an Al-contacted device (d~3.5 nm). The inset shows the dI/dV measurements taken at 2K with $V_{DS}$ (y-axis) and $V_G$ (x-axis) range of –4mV to 4mV and 9.5V to 10.05V respectively.

Another important phenomenon that we have observed upon PMMA passivation is that the ON-state conductance of the p-FETs typically increases by $\sim$ 50% or more (Fig. 3), accompanied by the hysteresis elimination. This may be attributed to the screening of the gate electric fields by the surrounding polar water molecules, resulting in less effective gating of the nanotubes. It is also possible that water removal at the Pd-SWNT contacts restores the work function of Pd to a slightly higher value than that when the water molecules are adsorbed at the contacts.

## Critical Aspects Regarding Diameter, OFF States and Ambipolar Behavior

It has been shown that when the dielectric thickness and channel length are aggressively scaled down, mid-gap SB nanotube FETs and ohmically Pd contacted SWNT-FETs with zero SB both exhibit high leak currents, low ON/OFF ratios, and ambipolar characteristics (10). These are undesired for CMOS applications. The situation becomes worse for larger diameter (small band gap) nanotubes. The ambipolar characteristics and high leak currents in the OFF-states are due to the tunneling currents through the thin SBs (SB width $\sim t_{ox}$) in the OFF-states. Here, our p- and n-FETs with d=2-3 nm SWNTs and $t_{ox}$~10 nm admittedly exhibit noticeable leak currents due to the thin oxide, and will show stronger ambipolar behavior when $t_{ox}$ is further reduced. Thus far, high ON current requires ohmic contacts that are accessible only with d>~2 nm tubes. Low OFF-current requires not only reasonably large band gaps, but also innovative device structure designs. This is an issue that begins to be tackled theoretically (11,12). Most recently, we have identified experimental approaches to solve this problem and obtain high ON and low OFF currents for SWNT-FETs. Detailed results will be presented separately.

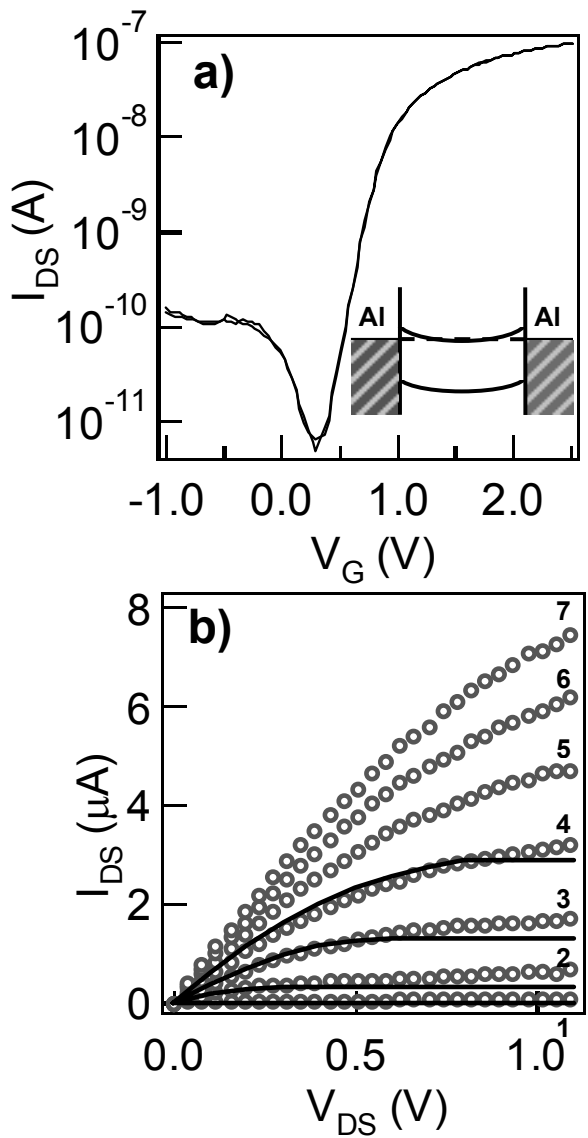

Fig. 4. Transport properties of an Al contacted SWNT n-FET with d ~ 2.8 nm and L ~ 2.5 μm. (a) Double-sweep $I_{DS}$-$V_G$ data for the device at $V_{DS}$ = 10 mV. (b) Output characteristic for the same nanotube FET. The circles are the experimental data while the solid lines represent the square law fit. Curves 1-7 correspond to $V_G$ = 0.7 – 2.5 V in 0.3 V steps.

### Threshold Voltages

Our n-FETs operate in the accumulation-mode while the p-FETs are normally in the ON-state at $V_G = 0$ (depletion-mode). The average threshold voltages for the n- and p-FETs (heavily p-doped Si as the back gate) are $V_{th} \sim 0.3$ and $0.6$ V respectively with $\sim \pm 0.5$ V of variation among different batches of samples. This variation is attributed to the difference in the band gap of individual nanotubes ($E_g \propto 1/d$) and a deviation in the fix charge density in the $SiO_2$ films. For circuit design, it is desirable to have both p- and n-FETs operating in the accumulation mode with small $V_{th}$. This can be achieved by tuning the gate material with proper work function. We are currently working towards achieving this goal.

### Complementary Logic

The ability to obtain high performance n- and p-FETs by changing the S/D metal type allows for simple demonstration of complementary nanotube circuits. Fig. 6a shows the $I_{DS}$-$V_G$ curves for the p- and n-FETs fabricated on the same chip by using Pd and Al contact electrodes respectively. Both devices show similar electrical properties with $R_{ON} \sim 35 - 40$ k$\Omega$. Transfer characteristic of a voltage-gain inverter, obtained by connecting these complementary FETs, is shown in Fig. 6b. This NOT gate demonstrates a new scheme of circuit design, which involves tuning the contacts rather than channel doping as a mean of controlling the properties of individual circuit elements.

### Conclusions

The new advancements in SWNT-FETs presented in this paper include (i) obtaining air-stable p- and n-FETs by varying S/D metals, and complementary inverters enabled by this approach. (ii) Demonstrating SB-free MOSFET-like operation of p-type SWNT transistors with Pd contacts. (iii) Attributing the transparency of the contacts not only to the work function of the metal contacts, but also to other important factors such as metal-tube interfacial chemical interactions. (iv) Eliminating hysteresis for accurate mobility and threshold voltage determination. (v) Observing high carrier mobility near 4000 $cm^2$/V.s for both electrons and holes and high ON-state current densities.

### Acknowledgement

We are grateful to Prof. Mark Lundstrom and Jing Guo for theoretical insights. This work was supported by MARCO MSD Focus Center, DARPA/Moletronics, SRC/AMD, the Stanford Initiative for Nanoscale Materials and Processes (INMP), a SRC Peter Verhofstadt Graduate Fellowship (A. J.), a Sloan Research Fellowship, Packard Fellowship and a Dreyfus Teacher-Scholar Award.

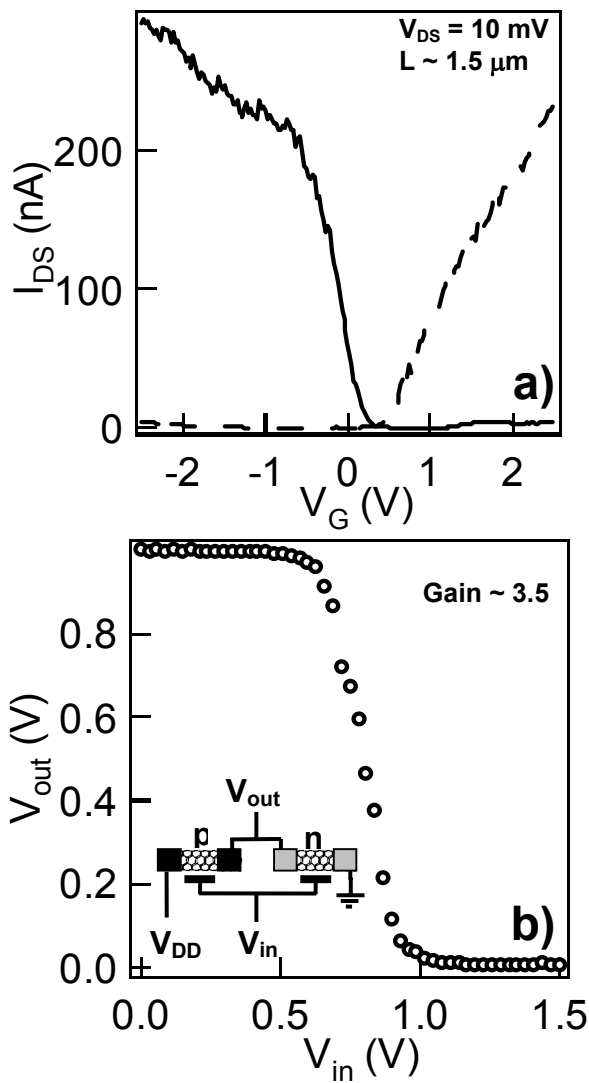

Fig.6. (a) Transfer characteristics for Pd-SWNT (solid) and Al-SWNT (dashed) FETs (L ~ 1.5 μm, d ~ 3 nm), fabricated on the same chip. (b) Transfer characteristics of a complementary voltage inverter obtained by using the two FETs in (a), showing a gain of ~ 3.5. The operating voltage is $V_{DD} = 1$V.